\documentstyle[prd,aps]{revtex}
\begin{document}
\draft
\preprint{WUGRAV-96-3}
\title{VACUUM FLUCTUATIONS AND COSMOLOGY\cite{LP}}
\author{L. P. Grishchuk}
\address{McDonnell Center for the Space Sciences, Department of Physics,\\
Washington University, St. Louis, Missouri 63130, U.S.A.\\
and\\
Sternberg Astronomical Institute,\\
Moscow University, 119899 Moscow V-234, Russia}
\date{To be published in Proceedings of the Les Houches Summer School, 
1995\\ ``Quantum Fluctuations''}
\twocolumn[
\maketitle
\widetext
\begin{abstract}
The notion of vacuum fluctuations of the gravitational field
plays important role in cosmology. The strong variable gravitational
field of the very early Universe amplifies these fluctuations and
transforms them into macroscopical cosmological perturbations.
Since the underlying process is the parametric amplification,
the perturbations are placed in squeezed vacuum quantum states.
It is possible that we already see the manifestations of these
processes in certain cosmological observations.  
\end{abstract}
\pacs{Preprint No. WUGRAV-96-3}
\twocolumn
]
\narrowtext
\section{Electrodynamics and Gravidynamics}

The concept of the zero-point (vacuum) quantum fluctuations is best known 
in the context of theoretical and experimental studies of the electromagnetic 
field. Even in this case, there still remain delicate questions about the 
meaning of vacuum fluctuations and whether or not they produce, like other 
forms of matter, their own gravitational field. Needless to say, vacuum 
fluctuations of the gravitational field itself seem to be even more enigmatic. 
The very quantization of the gravitational field is often considered as 
something mysterious. One will often be reminded that a fully consistent 
theory of quantum gravity is still a fundamental unsolved problem. 
Nevertheless, there are some reasons to believe that we may be facing the 
manifestations of the gravitational vacuum fluctuations already now, 
in cosmological observations. To make our point clearer, we will start from 
the presentation of the gravitational theory~---~the Einstein's general 
relativity~---~in the form similar to the theory of the electromagnetic field.

The laws of electrostatics are expressed by the Poisson equation for
the electrostatic potential $\phi$: 
\begin{equation}
  \Delta\phi = -4\pi\rho
\end{equation}
where $\rho$ is the charge density and $\Delta$ is the Laplace operator, 
$\Delta = {\partial^2 \over \partial x^2} 
+ {\partial^2 \over \partial y^2} 
+ {\partial^2 \over \partial z^2}.$
On the route to electrodynamics, one replaces 1 electrostatic 
potential $\phi$ by 4 electrodynamic potentials $A^\mu$ and the 
1-component source $\rho$ by the 4-component source $j^\mu$ 
(Greek indices $\mu ,\nu \ldots$ run from 0 to 3). One achieves the
relativistic generalization of the theory by replacing the Laplace 
operator $\Delta$ by the D'Alambert operator $\Box$,
$\Box = -{1\over c^2} {\partial^2\over \partial t^2}+\Delta$,
thus making the electromagnetic waves possible. As a result, one arrives 
at the equations  of electrodynamics, the Maxwell equations:
\[
\Box A^\mu = -{4\pi \over c} j^\mu
\]
or, in other notations,
\begin{equation}
  {{A^\mu}_{,\alpha}}^{,\alpha} = {4\pi \over c} j^\mu \quad .
\end{equation}
The Maxwell equations written in the form (2) assume that the gauge 
condition ${A^\alpha}_{,\alpha} = 0$ is chosen. Otherwise, without fixing
a gauge, the general form of the equations is
\begin{equation}
  {{A^\mu}_{,\alpha}}^{,\alpha} - {{A^\alpha}_{,\alpha}}^{,\mu}
= {4\pi \over c} j^\mu \quad .
\end{equation}

This form of equations makes the conservation law 
${j^\mu}_{,\mu} = 0$ 
automatically satisfied as a consequence of the equations.

The 4-vector potential $A^\mu$ and the 4-vector current $j^\mu$ are
defined	in the Minkowski space-time 
\begin{equation}
   d\sigma^2 = \eta_{\alpha\beta}~dx^\alpha ~dx^\beta
             = c^2 dt^2 - dx^2 - dy^2 - dz^2 \quad ,
\end{equation}
they are functions  of the Lorentzian coordinates $t,~x,~y,~z$. 
If for some reason one needs to introduce arbitrary curvilinear coordinates 
in the Minkowski space-time,
\begin{equation}
 d\sigma^2 =\gamma_{\alpha\beta} ~dx^\alpha ~dx^\beta \quad ,
\end{equation}
one is free to do this. Under arbitrary transformation of coordinates,
the objects $A^\mu$ and $j^\mu$ transform  as components of the 4-vectors. 
In Eq.~(3), one will need to replace ordinary derivatives by covariant 
derivatives defined with respect to the flat space-time metric (5) written 
in curvilinear coordinates. 

Being guided by the transition from electrostatics to electrodynamics, 
we can now try to guess how might look like the equations of gravidynamics, 
that is the relativistic field-theoretical generalization of the Newton law. 

The gravistatics is described by the equation similar to Eq.~(1): 
\[
    \Delta\phi = 4\pi G\rho
\]
where $\phi$ is the gravitational potential, and $\rho$ is the mass density. 
Since the mass density $\rho$ is only one component of the 10-component 
quantity, the energy-momentum tensor $\tau^{\mu\nu}$, one expects that
the source term of the relativistic gravitational field will be  
the full energy-momentum tensor $\tau^{\mu\nu}$. Accordingly, one needs
to replace 1 Newtonian potential $\phi$ by 10 gravidynamic potentials 
$h^{\mu\nu}$. The 10 components of $h^{\mu\nu}$ comprise, similarly 
to $\tau^{\mu\nu}$, a symmetric second rank tensor. To make the theory 
relativistic and to allow for a finite velocity of propagation and waves, 
one needs to replace the elliptic differential (Laplace) operator $\Delta$
by the hyperbolic differential (D'Alambert) operator $\Box$. And finally, 
being aware of the universal character of the gravitational field, 
one expects that the energy-momentum tensor $t^{\mu\nu}$ of the 
gravitational field itself should participate on the equal
footing with the $\tau^{\mu\nu}$ as the additional source term in 
the gravitational equations, thus making the equations nonlinear. 
One can guess that the equations of gravidynamics should have the form
\[
 \Box h^{\mu\nu} = -{16\pi G\over c^4} (t^{\mu\nu} + t^{\mu\nu})\quad .
\]
By analogy with the transition from electrostatics to electrodynamics, 
one may expect that this form of equations assumes that the gauge 
condition ${h^{\mu\nu}}_{,\nu} =0$ is already chosen.
In arbitrary gauge, the general form of equations is likely to be
\FL
\begin{eqnarray}
&& { {h^{\mu\nu}}_{,\alpha}}^{,\alpha}
  -{ {h^{\mu\alpha}}_{,\alpha}}^{,\nu}
  -{ {h^{\nu\alpha}}_{,\alpha}}^{,\mu}
  +\eta^{\mu\nu}{h^{\alpha\beta}}_{,\alpha ,\beta}\nonumber\\
&=& {16\pi G\over c^4} \bigl( t^{\mu\nu} + \tau^{\mu\nu}\bigr) \quad . 
\end{eqnarray}
This form of equations guarantees that the conservation law
$(t^{\mu\nu}+\tau^{\mu\nu})_{,\nu}=0$ is a direct consequence of the 
equations. It is obtained by taking derivatives from both sides of Eq.~(6).

The quantities $h^{\mu\nu}$, $\tau^{\mu\nu}$, and $t^{\mu\nu}$ are
assumed to be given in the Minkowski space-time (4). They are functions 
of the Lorentzian coordinates $t,~x,~y,~z$. However, one is free to 
introduce arbitrary curvilinear coordinates in the Minkowski space-time 
and to write the flat space-time metric in the form (5). Under arbitrary
coordinate transformation, the quantities 
$h^{\mu\nu}$, $\tau^{\mu\nu}$, $t^{\mu\nu}$
transform as symmetric second rank tensors. To write Eq.~(6) in arbitrary 
curvilinear coordinates, one needs to replace ordinary derivatives (coma's) 
by covariant derivatives (semicolon's) defined with respect to the 
metric (5). As a result, one obtains the equations 
\FL
\begin{eqnarray}
&& { {h^{\mu\nu}}_{;\alpha}}^{;\alpha}
  -{ {h^{\mu\alpha}}_{;\alpha}}^{;\nu}
  -{ {h^{\nu\alpha}}_{;\alpha}}^{;\mu}
  +{\gamma^{\mu\nu}h^{\alpha\beta}}_{;\alpha ;\beta}\nonumber\\
&=&{16\pi G\over c^4} \bigl( t^{\mu\nu}+\tau^{\mu\nu}\bigr) \quad .
\end{eqnarray}

Using the variational principle, Eqs.~(7) can be derived from the 
action $S$ consisting of the gravitational part $S^g$ and the matter
part (including interaction with gravity) $S^m$:
\[
  S = S^g+S^m , ~~
  S^g = -{c^3\over 16\pi G}\int d^4x L^g , ~~
  S^m = {1\over c} \int d^4x L^m ~ .
\]
Specific Lagrangians $L^g,~L^m$ determine, of course, 
the entire content of the theory, but we will not need their explicit form 
in our brief discussion. The energy-momentum tensors
$t^{\mu\nu}$, $\tau^{\mu\nu}$
are defined in a traditional field-theoretical manner, as variational 
derivatives of Lagrangians with respect to the 
metric tensor $\gamma_{\mu\nu}$. The energy-momentum tensor of the
gravitational field is 
$t_{\mu\nu} =-{c^4\over 8\pi G}~{1\over \sqrt{-\gamma}}
{\delta L^g\over \delta\gamma^{\mu\nu}}$,
the energy-momentum tensor of the matter sources (including their 
interaction with gravity) is
$\tau_{\mu\nu} ={2\over\sqrt{-\gamma}}
{\delta L^m\over \delta\gamma^{\mu\nu}}$.
Like in the case of electrodynamics, gravidynamics admits (in addition 
to the freedom of making arbitrary transformations of coordinates) 
the gauge freedom, that is a possibility of replacing the participating 
dynamical variables by new ones constructed from the old ones according 
to a certain prescription (gauge transformations). One can use this freedom 
for imposing gauge conditions, for instance, ${h^{\mu\nu}}_{;\nu}=0$.  

The point of the discussion above is that Eqs.~(7), (or Eqs.~(6) 
if we use the simplest, Lorentzian coordinates) which we envisaged 
on physical grounds, are exactly the Einstein equations, although written 
in a somewhat unfamiliar fashion. We have outlined the field-theoretical 
formulation of the general relativity, in contrast to the usual geometrical 
formulation. The link between the two formulations is realized by introducing 
the new functions 
$\sqrt{-g}~g^{\mu\nu}\equiv\sqrt{-\gamma}(\gamma^{\mu\nu}+h^{\mu\nu})$
and interpreting the functions $g_{\mu\nu}$ as the metric tensor of the
curved space-time: $ds^2=g_{\mu\nu}dx^\mu~dx^\nu$. 
The matter energy-momentum tensor $T^{\mu\nu}$ participating
in the geometrical Einstein equations is defined as the variational
derivative of the $L^m$ with respect to what is now metric, 
that is $g^{\mu\nu}$.  By using $g^{\mu\nu}$ and $T^{\mu\nu}$, 
one can transform Eqs.~(7) to the familiar geometrical representation 
of the Einstein equations:  
\[
  R_{\mu\nu} -{1\over 2} g_{\mu\nu} R
= {8\pi G\over c^4} T_{\mu\nu} \quad .
\]
The economy and beauty of the geometrical formulation, in which a 
single object $g_{\mu\nu}$ plays the double role of the metric tensor 
and the functions describing gravity, is accompanied by the loss of the
gravitational energy-momentum tensor $t^{\mu\nu}$ in exchange to ugly 
constructions called pseudotensors.

The two formulations of the general relativity are essentially two sides
of the same theory. The use of one or another framework becomes a
matter of taste. For our purposes, it will be convenient to think
of the general relativity in terms of the field-theoretical approach,
as of gravidynamics, similar in many aspects to electrodynamics. We will
discuss various cosmological models, but from the field-theoretical 
position, they are simply specific gravitational fields. 
For instance, the averaged structure of the observed part of the Universe 
is well approximated by a homogeneous isotropic cosmological model which 
can be described, in geometrical language, as curved space-time with 
the metric tensor
\begin{equation}
  ds^2 = g_{\mu\nu}~dx^\mu~ dx^\nu
= c^2dt^2 -a^2(t)(dx^2+dy^2+dz^2) \quad .
\end{equation}
The function $a(t)$ is called cosmological scale factor. This function
shows how the distance between the remote cosmological objects varies 
with time. We can also think of this cosmological model as of a 
gravitational field determined by the gravitational potentials
\begin{equation}
  h_{00} = a^3(t)-1 \quad ,\qquad
  h_{11} = h_{22} = h_{33}= 1-a(t) \quad ,
\end{equation}
specified in flat space-time with the metric tensor (4). From this position,  
the function $a(t)$ describes the strength and variability of the
gravitational field. In our further discussion this function will play the
role of the pump field acting upon quantized perturbations of the cosmological
gravitational field and whatever matter sources present in our model.

\section{Amplification of Classical Cosmological Perturbations}

We happened to live in approximately homogeneous and isotropic Universe. 
The observed volume of space includes immense number of galaxies and various 
sorts of radiation. On scales of galaxies and their clusters, matter is 
concentrated in well discernible systems.  However, on larger scales 
accessible for observations, the distributions of matter, its velocity, 
and accompanying gravitational field do not show any significantly 
preferred places or directions. The  photons of the microwave 
background radiation traveled to us for 10 billion years and, yet, 
the temperature of the radiation visible in different directions 
on the sky is remarkably the same. {\it A priori}, one would not 
be very surprised if one side of the observed sky contained, 
say, 20\% more matter than the other. For instance, we know that our 
own Galaxy is a flattened system of stars in which the Sun is located 
not very far from the edge of the disk. And there would be nothing 
wrong in having temperature of the microwave background radiation 
in one direction on the sky, say, 20\% higher or lower than in others. 
However, this is not the case. On large scales, the deviations from 
homogeneity and isotropy are small. Most convincingly, this is demonstrated 
by the fact that the measured large angular scale anisotropies of the 
microwave background radiation temperature $\delta T/T$ have the level 
of only $10^{-5}-10^{-6}$. So, when we work with homogeneous 
isotropic cosmological models plus small perturbations, this is not just 
a mathematical simplification, this is a reflection of observational 
data about the real world.

There are three ways in which one can perturb a homogeneous and isotropic 
distribution of matter and fields. First, one can compress matter in
various places perturbing its mass density, velocity and the accompanying
gravitational field. This is called cosmological density perturbations.
Second, one can provide the elements of matter with small rotational 
velocities (without perturbing the mass density) which will also be
accompanied by perturbations of gravitational field. This is called 
cosmological rotational perturbations. And third, one can perturb
gravitational field itself. This is called cosmological gravitational waves.
As we know, the equations of gravidynamics allow gravitational waves. In
contrast to gravitational waves emitted by localized sources, we will be
dealing with waves as excitations of gravitational field in the entire 
Universe.

Let us concentrate on gravitational waves --- the entity analogous 
to electromagnetic waves. 

The gravitational field of an idealized, strictly homogeneous 
and isotropic universe, is described by the potentials (9). We want to
consider a more realistic universe, which includes gravity-wave 
perturbations. The potentials can now be written in the form 
$h^{\mu\nu}=h^{\mu\nu (0)} + h^{\mu\nu (1)}$
where  $h^{\mu\nu (0)}$
is the zero-order approximation (9) and 
$h^{\mu\nu (1)}$ 
are the gravity-wave corrections. Since we expect the perturbations 
to be small, we can neglect quadratic and higher-order terms of
$h^{\mu\nu (1)}$ 
in Eq.~(6). By using the available gauge freedom we can eliminate the 
components
$h^{oo(1)}$, $h^{oi(1)}$
(Latin indices $i,j,\ldots$ run from 1 to 3). The remaining components          
$h^{ij(1)}$
are functions of $t,~x,~y,~z$ and can be Fourier decomposed over spatial 
harmonics 
$e^{i{\bf nx}}$, 
$e^{-i{\bf nx}}$, 
where ${\bf n} =(n^1,n^2,n^3)$
is arbitrary wavevector and the wavenumber $n$ is 
$n=({n^1}^2 + {n^2}^2 + {n^3}^2)^{1\over 2}$.  

Let us write down one of the Fourier components:  
\begin{equation}
      h_{ij}^{(1)} (t,{\bf x})
     = \sum_{s=1}^2 {\stackrel{s}{p}}_{ij}({\bf n})
\left[ {\stackrel{s}{\mu}}_n(t) e^{i{\bf nx}}
     + {\stackrel{s}{\mu^\ast}}_n(t)e^{-i{\bf nx}} \right] \quad .
\end{equation}
The two polarization tensors ${\stackrel{s}{p}}_{ij}$ $(s = 1, 2)$ 
satisfy the conditions
\[
 {\stackrel{s}{p}}_{ij}n^j = 0, ~~
 {\stackrel{s}{p}}_{ij}\delta^{ij} = 0, ~~
 {\stackrel{s}{p}}_{ij}
 {\stackrel{s'}{p}}^{ij} = 2\delta_{ss'}, ~~
 {\stackrel{s}{p}}_{ij}(-{\bf n}) = {\stackrel{s}{p}}_{ij}({\bf n}).
\]
Gravitational waves have two independent polarization states.  

We should now derive the equation governing the time dependent
complex amplitudes 
${\stackrel{s}{\mu}}_n(t)$.
We should use Eqs.~(6) retaining in them only terms linear in
${h_{ij}}^{(1)}$. The nonlinear right-hand side of the equations 
will provide us with cross terms, products of the zero-order 
terms $h^{\mu\nu (0)}$ and the first-order terms 
${h_{ij}}^{(1)}$. These cross terms signify the interaction 
of the gravitational waves with the pump field. As a result,
we obtain the equation 
\begin{equation}
 \ddot{\mu}_n + n^2\mu_n
= \left[ {n^2(a^2-1)\over a^2} + {\ddot{a} \over a}
+ \Bigl( {\dot{a} \over a} \Bigr)^2\right]
   \mu_n -{\dot{a} \over a} \dot{\mu}_n \quad .
\end{equation}
Both polarization amplitudes ${\stackrel{s}{\mu}}_n(t)$ satisfy this 
equation. 

We can now introduce a new time parameter $\eta$ according to
the relationship $cdt=ad\eta$ and denote $d/d\eta$ by a prime. 
Equation~(11) acquires the form
\begin{equation}
  \mu_n^{\prime\prime}+\Bigl[ n^2-{a^{\prime\prime}\over a}\Bigr]
  \mu_n = 0 \quad .
\end{equation}
We have derived this equation using the field-theoretical approach, but this 
is exactly the same equation which could be derived (and was in fact 
originally derived) using the geometrical approach. In terms of $\eta$-time 
the metric (8)  is manifestly conformally-Minkowskian:
\[
  ds^2 = a^2(\eta )(d\eta^2 - dx^2 -dy^2 -dz^2 )\quad .
\]

The form of the equation (12) explains why certain phenomena in
cosmology and quantum optics are linked by common underlying physics.
Eq.~(12) can be interpreted as equation for a parametrically excited 
oscillator. The term $n^2-a^{\prime\prime}/a$ can be considered as square 
of the time dependent frequency of the oscillator (even though in the most 
interesting regime this term is negative). This equation can also be
treated as a kind of the Schr\"odinger equation for a particle with energy 
$n^2$ in the presence of the potential barrier 
$U(\eta)=a^{\prime\prime}/a$. 
Here we will mostly discuss gravitational waves, but other cosmological 
perturbations (rotational perturbations and density perturbations) 
are described by equations of similar nature and many portions of our 
discussion apply to them as well. 

Let us show how the gravitational pump field can amplify classical
gravitational waves, leading essentially to the production of standing 
gravitational waves. We will illustrate this process on a specific
example. This example is a currently popular cosmological model which  
may have some relevance to the real Universe.  

Let us assume that the function $a(\eta )$ has the following behavior
\begin{equation}
a(\eta) =
\cases{ a_i(\eta )= -{c\over H\eta} \qquad , 
        &$-\infty <\eta\leq\eta_1$\cr
        a_e(\eta )= {c(\eta -2\eta_1) \over H\eta_1^2} \quad ,
        &$~~~\eta \geq \eta_1$ \quad ,\cr
      }
\end{equation}
where $H$ is a constant, the $c/H$ has dimension of length. 
The $a(\eta )$ and its first time derivative 
$a^\prime (\eta )$ are continuous functions at the point 
$\eta =\eta_1$. 
In cosmology, function (13) describes the so called De Sitter era 
with the scale factor $a_i(\eta )$ followed, after the moment of 
time $\eta =\eta_1$, $\eta_1 < 0$, by the radiation-dominated era
with the scale factor $a_e(\eta )$. The potential barrier $U(\eta)$ 
in Eq.~(12) vanishes asymptotically for 
$\eta \rightarrow -\infty$, reaches the maximum 
at $\eta =\eta_1$, and drops to zero after the moment $\eta =\eta_1$.

Equation (12) admits simple exact solutions in the both eras. For
$\eta\leq \eta_1$, one has 
\begin{equation}
  \mu_i = A\left[ \cos (n\eta + \phi )
        - {1\over n\eta}\sin (n\eta + \phi)\right] \quad,
\end{equation}
where $A$ and $\phi$, $0\leq\phi <2\pi$ 
are arbitrary constants. In the asymptotic regime 
$\eta \rightarrow -\infty$, we have usual harmonic oscillations   
\begin{equation}
   \mu_i\approx A~\cos (n\eta + \phi ) \quad .
\end{equation}
For $\eta \geq \eta_1$, one has
\begin{equation}
  \mu_e = B~\sin (n\eta + \chi ) \quad ,
\end{equation}
where $B$ and $\chi$ are also arbitrary constants. 

Solutions (14), (16) and their first time derivatives should join 
continuously at $\eta =\eta_1$. This allows us to express the final 
amplitude $B$ and phase  $\chi$ in terms of the initial amplitude $A$ 
and phase $\phi$. It is convenient to introduce the notation 
$x = n\eta_1$. Then we have
\begin{equation}
  B^2(\phi )= {A^2\over 2x^4} [1+2x^4-\cos (2\phi +2x-2\alpha )] \quad ,
\end{equation}
where $\tan 2\alpha = {2x\over 1-2x^2}$, and
\begin{equation}
  \tan\chi (\phi )
= \sqrt{ {1+2x^4-\cos (2x+2\alpha )
         \over 1+2x^4+\cos (2x+2\alpha)}}~~
  {\sin (\phi + \beta_1) \over \sin (\phi +\beta_2)} \quad ,
\end{equation}
where
\[
  \tan\beta_1 = {\sin^2x-x\over\sin x\cos x -x}~~,\quad
  \tan\beta_2 = {\sin x\cos x-x\over\cos^2 x -x^2}~~.
\]

The parametric amplification is a phase sensitive process,
the numerical value of $B$ depends on the initial phase $\phi$. 
However, the root mean squared (r.m.s.)  final amplitude is always larger 
than $A$, that is a ``typical'' oscillation will always be amplified. 
To show this, integrate $B^2(\phi )$ 
by $\phi$ from 0 to $2\pi$, divide the result by $2\pi$, 
and take the square root of this number. You will get
\[
  \sqrt{\overline{B^2}} = A\sqrt{1+ {1\over 2x^4}} \quad .
\]
Since $|x| \ll 1$ for the frequencies $n$ of our interest, we obtain
\[
  \sqrt{\overline{B^2}} \approx {A\over \sqrt{2}~x^2} \gg A \quad .
\]
We can also calculate the variance of the amplitude and find out that the
variance is large. Namely,   
\[
        { \overline{B^4} - (\overline{B^2})^2 \over A^4 }
\approx {1\over 2}~ {(\overline{B^2})^2 \over A^4} 
\approx {1\over 8x^8} \gg 1 \quad ,
\]
that is the dispersion of the quadratic amplitude $B^2$
is of the same order as the mean value of this quantity.

Under the condition $|x| \ll 1$, the final phase $\chi$, as a function 
of the initial phase $\phi$, will be concentrated near two values of 
$\chi$ separated by $\pi$. From Eq.~(18) we have approximately
\[
   \tan\chi \approx -2x
   {\sin (\phi + \beta_1 ) \over \sin (\phi + \beta_2)} \quad .
\]
The function $\tan\chi$ stays constant and almost zero for all $\phi$
except for the values of $\phi$ near $\phi =0,~\pi ,~2\pi,\ldots$, 
where $\tan\chi$ diverges and changes sign. Almost all initial phases 
$\phi$ from the interval between 0 and $\pi$ produce (``squeeze'' to) 
the final phase $\chi_1\approx 0$, and almost all initial phases 
$\phi$ from the interval between $\pi$ and $2\pi$ produce (``squeeze'' to) 
the final phase $\chi_2 \approx \pi$.  

So far, we have discussed the time dependence of the Fourier
component (10). Let us now take into account the spatial dependence too.  
Consider, for simplicity, only one polarization state and write
\begin{equation}
  h_{ij}^{(1)} 
= p_{ij} (\mu e^{i{\bf nx}}+\mu^\ast e^{-i{\bf nx}})\quad .
\end{equation}
At the early stage $\eta\rightarrow -\infty$, take, in agreement with 
Eq.~(15),
\begin{equation}
  \mu\approx A~e^{-in\eta} \quad ,
\end{equation}
where $A=|A|e^{-i\phi}$. The component (19) can be rearranged to read
\[
  h_{ij}^{(1)} \approx 2|A|p_{ij}~
  \cos (n\eta - {\bf nx} + \phi ) \quad ,
\]
that is we have initially a traveling (not standing) wave propagating 
in the direction ${\bf n}$. In the limit 
$\eta \rightarrow +\infty$, the solution (20) will have the form
\begin{equation}
  \mu \approx B_1~e^{-in\eta} + B_2~e^{in\eta} \quad .
\end{equation}
Since we are discussing the asymptotics of the same solution, the 
Wronskian 
$\mu\mu^{\ast\prime} - \mu^\ast\mu^\prime = {\rm const}$ 
should be the same constant for 
$\eta\rightarrow -\infty$ 
and $\eta\rightarrow +\infty$.
This gives:                                
\begin{equation}
  |A|^2 = |B_1|^2 - |B_2|^2 \quad .
\end{equation}
>From the point of view of physics, this condition expresses the 
conservation law of the linear momentum. As we know, a ``typical'' 
final amplitude is much larger than the initial one, that is,
$\overline{|B_1|^2} \gg |A|^2$
and 
$\overline{|B_2|^2} \gg |A|^2$.
This means, that, as a rule, the number on the left hand side of 
Eq.~(22) is a result of almost complete cancellation of the two much 
bigger numbers on the right hand side of that equation. This allows 
us to neglect $|A|^2$ and to write 
\begin{equation}
    B_1 \approx |B| e^{-i\psi_1} \quad , \qquad
    B_2 \approx |B| e^{i\psi_2} \quad , 
\end{equation}
Using these values of $B_1$, $B_2$ in Eq.~(21), we can rearrange 
Eq.~(19) to read, in the limit $\eta\rightarrow +\infty$, 
\begin{equation}
  h_{ij}^{(1)} \approx 4|B|p_{ij}~
  \cos \left(n\eta + {\psi_1+\psi_2 \over 2} \right)
  \cos \left( {\bf nx} + {\psi_2-\psi_1 \over 2} \right)\quad .
\end{equation}

This formula demonstrates explicitly that, after the amplification 
process, we are dealing with a standing wave. Since we have omitted 
the term $|A|^2$ in Eq.~(22), neglecting in this way the initial 
momentum, expression (24) is a strict standing wave. 
Acting more accurately, we would get an almost standing wave. 
The amplified initial traveling wave combines with the generated 
wave traveling in the opposite direction to form an (almost) 
standing wave. If we have had a strict standing wave initially, 
it would be amplified (except in the very narrow interval of 
inappropriate initial phases) maintaining its standing wave pattern. 

We have prepared ourselves for quantum-mechanical treatment of
this problem. Classical process of amplification requires a nonzero  
initial amplitude. Otherwise, the final amplitude will also be zero.
Quantum-mechanically, we do not need initial amplitude, or, we can say, 
some minimal level of the amplitude is provided by the zero-point 
quantum fluctuations. Since we can think of a vacuum state as of 
a set of fluctuations with all phases, we can expect that the
amplification process will necessarily take place. 
Quantum-mechanically, this process describes creation of strongly 
correlated pairs of waves (particles) by the external pump field.  
The already briefly discussed distributions of the final amplitudes 
and phases (assuming that the initial phase $\phi$ is distributed 
randomly and evenly in the interval $0-2\pi$) are related to the 
phenomenon of squeezing. 

\section{Quantized Cosmological Perturbations and Squeezing}

It is now clear what we are going to quantize --- the
fluctuating part $h^{\mu\nu(1)}$ of the gravitational field.
If you think of gravitation as of geometry, you may be lead to 
the idea of quantizing, emitting and absorbing, pieces of space 
and intervals of time. If you think of gravity as of a field, 
you are lead to the idea of quantizing the field in no more 
mysterious fashion (even if more difficult technically) than 
quantization of the electromagnetic field. There is not much 
sense in quantizing the pump field $h^{\mu\nu (0)}$, we are not 
going to discuss such issues as creation of universes, we are 
going to discuss creation of excitations in the Universe. 
Our theory of small excitations is essentially linear. 
The fundamental nonlinearity of gravidynamics provides us with 
the parametric interaction of the quantized perturbations with 
the pump filed. In contrast to quantum optics, we do not need 
any intervening nonlinear optical medium in order to couple the 
quantized part of the field with the pump field, gravity is 
inherently nonlinear.     

Let us write the gravity-wave field operator in the general form
\FL
\begin{eqnarray}
       h_{ij}(\eta ,{\bf x})
    &=& {C\over a(\eta )}~{1\over (2\pi )^{3/2}}
       \int_{-\infty}^\infty d^3{\bf n} \sum_{s=1}^2
       {\stackrel{s}{p}}_{ij}({\bf n}) {1\over \sqrt{2n}}\nonumber\\
&\times & \left[ {\stackrel{s}{c}}_{\bf n} (\eta ) e^{i{\bf nx}} 
      +{\stackrel{s}{c}}_{\bf n}^{\dag} (\eta )e^{-i{\bf nx}} \right] \quad .
\end{eqnarray}
Being armed with the gravitational energy-momentum tensor 
$t^{\mu\nu}$, 
we can find the normalization factor $C$. In the traditional manner, 
we require that the vacuum energy of each mode be equal to a 
``half of the quantum''.  From this requirement we find 
$C=\sqrt{16\pi}~l_{Pl}$ where $l_{Pl}$ is the Planck length, 
$l_{Pl}=(G\hbar /c^3)^{1\over 2}$. 
The gravitational constant $G$ and velocity of light $c$ participate 
in the definition of
$t^{\mu\nu}$.
The Planck constant $\hbar$ enters because of our normalization. All 
the three fundamental constants combine in $l_{Pl}$ and appear in $C$, 
explicitly demonstrating that we are working with quantum gravity.

The creation and annihilation operators 
$c_{\bf n}(\eta )$,
$c_{\bf n}^{\dag} (\eta )$
for each polarization state are governed by the Heisenberg equations 
of motion
\begin{equation}
     {d c_{\bf n}(\eta ) \over d\eta}
   = -i [c_{\bf n}(\eta ),H] \quad ,\qquad
     {d c_{\bf n}^{\dag}(\eta ) \over d\eta}
   = -i [c_{\bf n}^{\dag} (\eta ),H] 
\end{equation}
where $H$ is the Hamiltonian for the modes involved. This Hamiltonian
follows from the general Lagrangian of our theory, truncated at the
appropriate order.  Combining those formulas with the understanding 
of what kind of processes we are dealing with, one can derive 
\[
   H = nc_{\bf n}^{\dag} c_{\bf n}
     + nc_{-{\bf n}}^{\dag} c_{-{\bf n}}
     + 2\sigma (\eta ) c_{\bf n}^{\dag} c_{-{\bf n}}^{\dag}
     + 2\sigma^\ast (\eta ) c_{\bf n} c_{-{\bf n}} \quad .
\]
The coupling function $\sigma (\eta )$ is determined by the pump 
field:
$\sigma (\eta )={i\over 2} {a' \over a}$.

The form of the Hamiltonian dictates the form of the solution
(Bogoliubov transformation) to Eq.~(26):
\begin{eqnarray}
   c_{\bf n}(\eta ) =
&& u_n(\eta )c_{\bf n} (0) 
   + v_n(\eta )c_{-{\bf n}}^{\dag} (0) \quad , \nonumber\\
   c_{\bf n}^{\dag}(\eta ) =
&& u_n^\ast (\eta )c_{\bf n}^{\dag}(0) 
   + v_n^\ast (\eta )c_{-{\bf n}} (0) \quad . 
\end{eqnarray}
The initial values 
$c_{\bf n}(0)$, $c_{\bf n}^{\dag} (0)$  
of the operators are taken at the very early times,
$\eta \rightarrow -\infty$, 
long before the interaction became effective. The two complex functions 
$u_n(\eta )$, $v_n(\eta )$
satisfy the equations
\begin{equation}
  i{du_n \over d\eta} = nu_n + i{a' \over a} v_n^\ast \quad ,\qquad
  i{dv_n \over d\eta} = nv_n + i{a' \over a} u_n^\ast 
\end{equation}
and the condition
\begin{equation}
  |u_n|^2 - |v_n|^2 = 1
\end{equation}
which guarantees that the commutator relationship 
$[c_{\bf n}(\eta ),c_{{\bf n}'}^{\dag}(\eta )]=\delta^3({\bf n}-{\bf n}')$
is satisfied for all times. The condition (29) is automatically fulfilled  
if one introduces the (squeeze) parameters $r$, $\theta$, $\phi$ 
according to the definition
\[
   u = e^{i\theta} \cosh~r \quad , \qquad
   v = e^{-i(\theta -2\phi )} \sinh~r \quad .
\]
In our problem, these parameters are functions of time.

Using Eqs. (27) one can present the field operator (25) in the form
\FL
\begin{eqnarray}
&& h_{ij} (\eta ,{\bf x})
= {C\over (2\pi )^{3/2}} \int_{-\infty}^\infty d^3{\bf n}
  \sum_{s=1}^2~{\stackrel{s}{p}}_{ij} ({\bf n})
   {1\over \sqrt{2n}}\nonumber\\
&\times & \left[ {\stackrel{s}{h}}_n (\eta ) e^{i{\bf nx}}~
                 {\stackrel{s}{c}}_{\bf n}(0)
                +{\stackrel{s}{h^\ast}}_n (\eta ) e^{-i{\bf nx}}~
                 {\stackrel{s}{c}}_{\bf n}^{\dag} (0) \right]~~,
\end{eqnarray}
where the functions ${\stackrel{s}{h}}_n(\eta )$ are
\[
  {\stackrel{s}{h}}_n(\eta ) = {1\over a(\eta )} 
  [{\stackrel{s}{u}}_n(\eta ) + v_n^{s\ast} (\eta )]
= {1\over a(\eta )} {\stackrel{s}{\mu}}_n(\eta ) \quad .
\]
>From equations (28) for
${\stackrel{s}{u}}_n(\eta )$,
${\stackrel{s}{v}}_n(\eta )$
we recover classical equations of motion (12) for the functions
${\stackrel{s}{\mu}}_n(\eta )$.

We have reduced our problem to the problem studied also in other areas
of physics. By cosmological reasons, we make the simplest assumption 
about the initial state, namely that each mode of the field
was initially in the vacuum state $|0_{\bf n}\rangle$ defined by
${\stackrel{s}{c}}_{\bf n}(0)|0_{\bf n}\rangle =0$.
In the Schr\"odinger picture, the initial vacuum state 
$|0_{\bf n}\rangle |0_{-{\bf n}}\rangle$
transforms into a two-mode squeezed vacuum quantum state. 
To see this formally, we can note that the Bogoliubov transformation
(27) can be presented as
\[
  c_{\bf n}(\eta ) = RS~c_{\bf n}(0)S^{\dag}R^{\dag}~, ~~
  c_{\bf n}^{\dag}(\eta ) = RS~c_{\bf n}^{\dag}(0)S^{\dag}R^{\dag}
\]
where
\begin{equation}
       S(r,\phi ) = \exp 
\left[ r\Bigl( e^{-2i\phi} c_{\bf n}(0)c_{-{\bf n}}(0)
       - e^{2i\phi} c_{\bf n}^{\dag}(0)c_{-{\bf n}}^{\dag}(0) \Bigr)
\right]
\end{equation}
is the two-mode squeeze operator, and
\[
        R(\theta ) = \exp 
\Bigl[ -i\theta 
\Bigl(  c_{\bf n}^{\dag}(0) c_{\bf n}(0)
      + c_{-{\bf n}}^{\dag}(0) c_{-{\bf n}}(0) \Bigr)\Bigr]
\]
is the two-mode rotation operator. In our case, the two modes are 
two waves traveling in the opposite directions. Creation of pairs of
correlated waves (particles) corresponds to classical picture of
generation of almost standing waves. 

We can also work with the one-mode squeezed states and associated 
standing waves. To reach this goal, introduce two new sorts of operators, 
${\stackrel{s}{a}}_{\bf n}$ and
${\stackrel{s}{b}}_{\bf n}$,
according to the rule
\[
       {\stackrel{s}{a}}_{\bf n} = {1\over \sqrt 2}
\Bigl( {\stackrel{s}{c}}_{\bf n} 
      +{\stackrel{s}{c}}_{-{\bf n}} \Bigr) \quad , \qquad
       {\stackrel{s}{b}}_{\bf n} = {1\over \sqrt 2}
\Bigl( {\stackrel{s}{c}}_{\bf n} 
      -{\stackrel{s}{c}}_{-{\bf n}} \Bigr) \quad . 
\]
In terms of these operators, the field (25) can be written as 
\FL
\begin{eqnarray}
&&  h_{ij}(\eta ,{\bf x}) = 
   {C\over \sqrt2 a(\eta )} {1\over (2\pi )^{3\over 2}}
   \int_{-\infty}^\infty d^3{\bf n} \sum_{s=1}^2
   {\stackrel{s}{p}}_{ij}({\bf n}) {1\over \sqrt{2n}}\nonumber\\
&\times &
   \Bigl[\Bigl( {\stackrel{s}{a}}_{\bf n}(\eta ) 
               +{\stackrel{s}{a}}_{\bf n}^{\dag}(\eta ) \Bigr)
                \cos {\bf nx} + 
\Bigl( {\stackrel{s}{b}}_{\bf n}(\eta ) 
      +{{\stackrel{s}{b}}_{\bf n}}^{\dag}(\eta ) \Bigr)
      \sin {\bf nx} \Bigr]\quad , \nonumber 
\end{eqnarray}
demonstrating the standing wave decomposition. Accordingly,  
the two-mode squeeze operator (31) factorizes into a product of two 
identical one-mode squeeze operators: 
\begin{eqnarray}
   S(r,\phi ) 
&=& \exp \Bigl[ {1\over 2} r\Bigl( e^{-2i\phi}~
   a_{\bf n}(0)a_{\bf n}(0) - e^{2i\phi}~
   a_{\bf n}^{\dag}(0)a_{\bf n}^{\dag}(0) \Bigr)\Bigr]\nonumber\\
&\times & \exp \Bigl[ {1\over 2} r\Bigl( e^{-2i\phi}~
   b_{\bf n}(0)b_{\bf n}(0) - e^{2i\phi}~
   b_{\bf n}^{\dag}(0)b_{\bf n}^{\dag}(0) \Bigr)\Bigr]. \nonumber
\end{eqnarray}
The ``particles'' described by $a_{\bf n}$ and $b_{\bf n}$ operators 
do already have zero linear momentum.  

We will work in the Heisenberg picture, in which the initial vacuum 
state does not change. It immediately follows from the representation 
(30) that the mean quantum-mechanical value of the field 
$h_{ij}(\eta ,{\bf x})$ is zero in every spatial point and at every 
moment of time: $\langle 0|h_{ij}(\eta ,{\bf x})|0\rangle =0$. 
As might have been expected, the variance of the field is not zero 
and does depend on time. Using a product of two expressions (30) 
one can show that
\begin{equation}
   \langle 0|h_{ij}(\eta ,{\bf x}) h^{ij}(\eta ,{\bf x}) |0\rangle 
=  {C^2 \over 2\pi^2} \int_0^\infty n \sum_{s=1}^2
   \Big| {\stackrel{s}{h}}_n(\eta )\Big|^2 ~dn \quad .
\end{equation}
The expression under the integral is usually called the power spectrum
of the field (in this case, the power spectrum of the quantity 
$h_{ij}h^{ij}$):
\begin{equation}
P(n) = {C^2 \over 2\pi^2} n \sum_{s=1}^2 
       \Big| {\stackrel{s}{h}}_n(\eta )\Big|^2 \quad .
\end{equation}
The integral (32) is always divergent in the ultra-violet 
limit $n\rightarrow \infty$, and sometimes divergent in the infra-red 
limit $n\rightarrow 0$. The ultra-violet divergence is caused by 
the high-frequency waves (above the barrier in Eq.~(12)) which 
were not affected by the amplification process and which survived 
from $\eta\rightarrow -\infty$ to $\eta\rightarrow +\infty$. 
The renormalization of Eq.~(32) (subtraction of the ``half of the 
quantum'' contributions from the final expression) eliminates the 
divergence and this part of the spectrum entirely, without changing 
substantially the lower-frequency part of the spectrum where significant 
amplification took place. As for the infra-red divergence, it arises 
if one assumes that the amplification process had been taking place for 
infinitely long time, which does not happen in practice.  So, 
in practice, the integral (32) has natural high-frequency and 
low-frequency cutoffs, and does not require renormalization.

For cosmological applications it is important to know the r.m.s. values 
of the field, its statistical properties, shape of the spectrum. 
Of course, they all are determined by the strength and variability 
of the pump field, and by the very nature of squeezed vacuum quantum 
states. In realistic cosmological models, the generated spectrum 
extends from wavelengths of order of a few meters up to the 
wavelenghts comparable with the Hubble radius $l_H \approx 10^{28}$~cm 
and beyond. Correspondingly, the squeeze parameter $r$ ranges 
from $r\approx 1$ at the short-wavelength end up to $r\approx 100$ 
and more for extremely long waves. The r.m.s. numerical values of the 
field are sufficiently high to believe that the quantum-mechanically 
generated cosmological perturbations may provide explanation to
some of cosmological observations. For instance, if the length parameter
$c/H$ in the simplest model (13) is 5 or 6 orders of magnitude larger than
$l_{Pl}$, the r.m.s. amplitude of the generated gravity-wave field is
appropriate to produce the large-angular-scale variations in the 
microwave background radiation at the level
$\delta T/T \approx 10^{-5}-10^{-6}$. 

The two most important features of strongly squeezed 
vacuum states --- small variances of phase and large variances 
of amplitude --- are responsible for the characteristic features 
of cosmological perturbations themselves and of the microwave background 
anisotropies caused by them. The small variances of phase and the 
associated standing wave pattern of the field make the spectrum (33) 
an oscillating function of frequency $n$ for a fixed moment of time 
$\eta$. The possibility of seeing these oscillations directly or
indirectly is one of the observational challenges of cosmology. 
On the other hand, large variances of amplitude should lead to large 
statistical deviations of the observable quantities from their expected 
values. One may recall that, in a squeezed vacuum state, the dispersion 
of the number operator $N$ is of the same order as the mean value of
$N$: 
$\langle N^2\rangle -\langle N\rangle^2 =2\langle N\rangle
(\langle N\rangle +1)$.
In the next section we will focus on the consequences of the large
variances of amplitude for statistical properties of the microwave
background anisotropies.  

\section{Statistics of the Microwave Background Anisotropies Caused
by Squeezed Cosmological Perturbations}

Why do we think that the quantum gravity processes in the very early
Universe may have something to do with current observations? 

The first point is that the mechanism studied above is universal, 
it does not require any hypothesis beyond the validity of general
relativity and basic principles of quantum field theory. 
One can believe that, in one or another amount, the quantum-mechanically 
generated perturbations should be present.  We can not make an exact 
prediction about the perturbations, because the pump field (the scale 
factor of the very early Universe) is not known. (And if it were known, 
there would be not too much to learn by testing our predictions.)   
But even the absence of perturbations on a certain level tells us 
something positive. By comparing theoretical predictions of various 
cosmological models with observations we restrict the possible behavior 
of the very early Universe rather than refine hypotheses on the 
generating mechanism.

The second point is that the measured large-angular-scale anisotropies 
are produced by cosmological perturbations with extremely long wavelengths. 
It is difficult to devise an alternative mechanism of their generation, 
especially if we do not want to make many additional assumptions. 
At the same time, the quantum-mechanical generation suggests itself 
as natural and unavoidable. 

If the observed large-angular-scale anisotropies are indeed caused 
by cosmological perturbations of quantum-mechanical origin 
(further theoretical study shows that they should be predominantly 
gravitational waves), the anisotropies should carry signatures 
of the underlying squeezed vacuum quantum states. One possibility
to prove (or disprove) the quantum-mechanical origin of the 
perturbations is to use their statistical properties.

Let us first discuss how anisotropies arise in the presence of the
gravitational field perturbations, regardless of what is the origin
of perturbations. In a strictly homogeneous and isotropic universe, 
the temperature of the microwave background radiation seen in all 
directions on the sky would be the same, $T$. The field (9) is 
independent of spatial coordinates and whatever its effect on 
propagating photons is, it does not depend on direction. 
In presence of additional gravity field potentials 
$h_{ij}(\eta ,{\bf x})$, 
the frequency of photons (and, hence, the temperature of the Planck 
distribution of the microwave radiation) changes along their paths. 
Even if the photons were emitted with exactly the same frequency 
(determined by the physical conditions of local equilibrium in points 
of their release) they will arrive to us from different directions 
with different frequencies, because they have traversed different 
paths in the field
$h_{ij}(\eta ,{\bf x})$. 
In a given direction on the sky, characterized by a unit vector
${\bf e}=(e^1,e^2,e^3)$, the actual measured temperature differs 
from its unperturbed value $T$. Using either geometrical or 
field-theoretical approach, one can show that the variation of 
temperature in a given direction ${\bf e}$ is expressed by the formula 
\begin{equation}
   {\delta T\over T}({\bf e}) = {1\over 2} \int_0^{w_1}
   {\partial h_{ij} \over \partial\eta} e^ie^j~d\eta \quad ,
\end{equation}
where $\partial h_{ij}/\partial\eta$ is taken along the integration 
path 
$x^i=e^iw$, $\eta =\eta_R-w$, from the event of reception $w=0$ 
to the event of emission $w=w_1=\eta_R-\eta_E$.  

In our case, the field $h_{ij}$ entering Eq.~(34) is a
quantum-mechanical operator (30). Hence, the $\delta T/T$ 
does also become an operator. Having defined in this way an observable 
and knowing the quantum state of the field $|0\rangle$, we can compute 
various expectation values.

Obviously, the mean value of $\delta T/T$ in every direction on the sky
is zero:
\[
   \langle 0| {\delta T\over T} ({\bf e}) |0\rangle = 0 \quad .
\]
The product of two operators 
${\delta T\over T} ({\bf e}_1)$ and
${\delta T\over T} ({\bf e}_2)$ for two different directions 
${\bf e}_1$ and ${\bf e}_2$ 
is a new operator.  The mean value of this operator is called 
the angular correlation function:
\begin{equation}
  K({\bf e}_1,{\bf e}_2)
= \langle 0|{\delta T\over T} ({\bf e}_1)
  {\delta T\over T} ({\bf e}_2) |0 \rangle  \quad .
\end{equation}
One can calculate this quantity by manipulating with the product of two
expressions (34) and performing integration over angular variables 
in the 3-dimensional wavevector ${\bf n}$ space. One can show, 
first, that the quantity 
$K({\bf e}_1,{\bf e}_2)$
depends only on the separation angle $\delta$ between the vectors
${\bf e}_1$, ${\bf e}_2$, not vectors themselves, and, second, 
that the quantity
$K({\bf e}_1,{\bf e}_2)$
takes on the form
\begin{equation}
    K({\bf e}_1,{\bf e}_2) = K(\delta )
= l_{Pl}^2 \sum_{l=l_{\rm min}}^\infty ~K_l P_l(\cos\delta )\quad .
\end{equation}
The coefficient $l_{Pl}^2$ is taken from $C^2$, other numerical 
coefficients are included in $K_l$. The $P_l(\cos\delta )$ 
are the Legendre polynomials. The lowest multipole $l_{\rm min}$ 
follows from the theory and, in case of gravitational waves, it is 
$l_{\rm min}=2$. The quantities $K_l$ are calculable from a given 
cosmological model plus perturbations, and they eventually depend 
on the pump field. For a zero separation angle $\delta =0$, 
one obtains the variance of ${\delta T\over T}({\bf e})$:  
\[
   \langle 0|{\delta T\over T} ({\bf e}) 
   {\delta T\over T} ({\bf e}) |0\rangle = K(0) 
  =l_{Pl}^2 \sum_{l=l_{\rm min}}^\infty K_l \quad .
\]

The numerical level of $\delta T/T$ is basically determined by the 
r.m.s. numerical level of the generated amplitudes of the perturbations. 
Large variances of the amplitude will be reflected in large statistical
deviations of the measured values of the observable 
${\delta T\over T}({\bf e}_1){\delta T\over T} ({\bf e}_2)$
from its expectation value (36). Let us calculate the variance 
$V({\bf e}_1,{\bf e}_2)$ of the operator
$v\equiv {\delta T\over T}({\bf e}_1){\delta T\over T}({\bf e}_2)$.
By definition, we have
\[
  V({\bf e}_1,{\bf e}_2) 
= \langle 0| v^2 |0\rangle - [\langle 0| v |0\rangle ]^2 \quad .
\]
This calculation requires us to deal with the product of 4 expressions 
(34). However, this calculation is easy to perform. One can show that,
first, the quantity $V({\bf e}_1,{\bf e}_2)$ depends only on the separation 
angle $\delta$, and, second, its concrete value is
\[
   V({\bf e}_1,{\bf e}_2)=V(\delta )=K^2(\delta )+ K^2(0)\quad .
\]
As could be expected, the standard deviation
\[
  \sigma (\delta ) = \sqrt{V(\delta )}
= \sqrt{K^2(\delta )+K^2(0)}
\]
is very big. Even for those separation angles for which the 
$K(\delta )$ vanishes, the standard deviation of the variable $v$
is as big as the mean value of this variable. If the anisotropies 
are indeed caused by the squeezed cosmological perturbations, 
this statistical feature must be present in the observational data. 
The problem is how to recognize it having only a single map of the sky. 
The measured anisotropies are essentially a single function of angular 
coordinates which assigns different numerical values of the temperature 
to different directions on the sky. 

A complication, specific for cosmology, is that we will always be
dealing with only one realization of our quantum state, or, 
in other words, with only one experiment, only one map.  
The calculated expectation values imply averaging over many 
realizations, many experiments, ``many universes'', but we have 
access to only one of them.  The meaningful comparison of the theory 
and experiment will require a kind of ergodic hypothesis allowing 
us to replace the quantum-mechanical expectation value of, say, 
observable $v$ by the procedure of averaging the measured 
${\delta T\over T}({\bf e}_1){\delta T\over T}({\bf e}_2)$
over many pairs of directions ${\bf e}_1$, ${\bf e}_2$ on a single map.

Even if it is known in advance that the anisotropies are induced by
squeezed perturbations, the extraction of cosmological information 
from the observations is not easy. The large statistical uncertainties 
discussed above make the discrimination of competing cosmological models 
(competing pump fields) a much more difficult problem than we wanted 
it to be. It is remarkable, however, that the existing and planned 
more accurate measurements can solve these problems and help us 
to learn how the Universe behaved when it was a tiny fraction of 
a second old. 
\acknowledgments
This work was supported by NASA grants NAGW 2902, 3874 and 
NSF grant 92-22902.

\begin{itemize}

\item Gravity-Wave Astronomy: Some Mathematical Aspects. In: {\it Current
Topics in Astrofundamental Physics}, Eds. N. Sanchez and A. Zichichi,
p. 435 (World Scientific, Singapore, 1992).

\item Cosmological Perturbations of Quantum-Mechanical Origin and
Anisotropy of the Microwave Background Radiation.  
In: \protect{\it Current Topics in Astrofundamental Physics, 
The Early Universe}, Eds. N. Sanchez and A. Zichichi, 
NATO ASI Series C, Vol. 467, p. 205
(Kluwer Academic Publishers, 1995).

\item Statistics of the Microwave Background Anisotropies Caused
by Cosmological Perturbations of Quantum-Mechanical Origin.
In: \protect{\it String Gravity and Physics at the Planck Scale}, 
Eds. N. Sanchez and A. Zichichi (Kluwer Academic Publishers, 1996)
(to appear); gr-qc/9511074.
\end{itemize}

\begin{references}
\bibitem[*]{LP}
More details on the subject of these lectures as well as references
to the original papers can be found in the Erice School contributions of
the author:
\end{references}
\end{document}